\documentclass[]{agujournal}

\journalname{Geophysical Research Letters}

\usepackage{placeins}

\begin{document}

------------------------------------------------------------------------ 

\title{Estimating Fault Friction from Seismic Signals in the Laboratory} 

\authors
{Bertrand Rouet-Leduc\affil{1}, Claudia Hulbert\affil{1}, David C. Bolton\affil{2}, Christopher X. Ren\affil{3}, Jacques Riviere\affil{2,4}, Chris Marone\affil{2}, Robert A. Guyer\affil{1}, Paul A. Johnson\affil{1}}

\affiliation{1}{Los Alamos National Laboratory, Geophysics Group, Los Alamos, New Mexico, USA}
\affiliation{2}{Department of Geosciences, Pennsylvania State University, University Park, Pennsylvania, USA}
\affiliation{3}{University of Cambridge, Department of Materials Science and Metallurgy, Cambridge CB3 0FS, UK}
\affiliation{4}{Institute of Earth Sciences (ISTerre), Grenoble Alpes University, CNRS, 38000 Grenoble, France}

\correspondingauthor{B. Rouet-Leduc}{bertrandrl@lanl.gov}

\begin{keypoints}
\item Machine learning models can discern the frictional state of a laboratory fault from the statistical characteristics of the seismic signal 
\item The use of machine learning uncovers a simple relation between fault frictional state and statistical characteristics of the seismic signal
\item The discovery of this equation of state also uncovers the hysterectic behavior of the laboratory fault
\item This equation of state between seismic signal power and friction generalizes to different stress conditions with the appropriate scaling
\end{keypoints}


\begin{abstract}
Nearly all aspects of earthquake rupture are controlled by the friction along the fault that progressively increases with tectonic forcing, but in general cannot be directly measured. We show that fault friction can be determined at any time, from the continuous seismic signal. 
In a classic laboratory experiment of repeating earthquakes, we find that the seismic signal follows a specific pattern with respect to fault friction, allowing us to determine the fault's position within its failure cycle.
Using machine learning, we show that instantaneous statistical characteristics of the seismic signal are a fingerprint of the fault zone shear stress and frictional state. Further analysis of this fingerprint leads to a simple equation of state quantitatively relating the seismic signal power and the friction on the fault. These results show that fault zone frictional characteristics and the state of stress in the surroundings of the fault can be inferred from seismic waves, at least in the laboratory.
\end{abstract}

\section*{Plain language summary}
In a laboratory setting that closely mimics Earth faulting, we show that the most important physical properties of a fault can be accurately estimated using machine learning to analyze the sound that the fault broadcasts. The artificial intelligence identifies telltale sounds that are characteristic of the physical state of the fault, and how close it is to failing. A fundamental relation between the sound emitted by the fault and  its physical state is thus revealed. 

\section{Introduction}

Most tectonic earthquakes take place when juxtaposed crustal blocks that are locked or slowly slipping overcome the static fault friction and abruptly slide past one another. A rupture initiates and propagates along the fault plane, eventually coming to a stop as the dynamic fault friction puts a brake on continued slip.  It is the frictional state that controls how the fault ruptures, its nucleation and how big the earthquake will ultimately become. The fault frictional state also controls when the next event may take place under a given tectonic (or anthropogenic) forcing (\cite{scholz2002mechanics,Marone1998}). 

Inferring the frictional state on faults, and where a fault is within its seismic cycle, is extremely challenging. Seismic wave recordings at the time of an earthquake can inform us about characteristics such as rupture velocity and can be used to calculate fundamental parameters such as earthquake magnitude (\cite{aki2002quantitative}), the evolution of elasticity following an earthquake (\cite{2007GeoRL..34.2305B,brenguier2008postseismic,curtis2006seismic,nakata2011near}) and slip distribution for instance (\cite{manighetti2005evidence}). However, seismic waves have not been used to directly examine the frictional state throughout the entire seismic cycle, nor its distribution along the fault. In fact, no geophysical data set has enabled the direct and continuous quantification of the fault frictional state. 

Frictional characteristics are determined primarily from theory, simulations and laboratory experiments 
(\cite{Rabinowicz1956,Fineberg2004,Dorostkar2017,scholz1968microfracturing,scholz2002mechanics,JGRB:JGRB51267,mclaskey2011micromechanics,MORGAN1997209,Madariaga2016,Kaproth2013}). Large scale stress simulations based on plate movements can provide estimates of stress and frictional state on a fault, but within significant error bounds (\cite{zoback1991tectonic,townend2013}). Computer models, including state-of-the-art simulations can be powerful but currently fall short in regards to predicting actual fault behavior. Nonetheless, simulations of the complex behavior of faulting are improving rapidly (\cite{Richards-Dinger983}) and laboratory experiments provide tremendous insight into frictional processes (\cite{scholz1968microfracturing,scholz2002mechanics,JGRB:JGRB51267,mclaskey2011micromechanics,brantut2008high}). Laboratory shearing experiments, involving an apparatus identical to that which produced the data that we analyze here, have been instrumental in the development of rate and state friction laws (\cite{scholz1998,Marone1998}).

In laboratory shear experiments that use fault blocks separated by fault gouge, many slip behaviors that resemble those observed in Earth can be induced, including stick-slip and slow-slip (\cite{Kaproth2013,Scuderi2016,mclaskey2011micromechanics,GRL:GRL27642}).  In particular, the fundamental Gutenberg-Richter relation for laboratory events (\cite{johnson2013}) is very similar to small-scale earth observations such as in mines (\cite{GRL:GRL25912}), tectonic regions (\cite{GJI:GJI5595}), and to the whole Earth (\cite{GJI:GJI135}), showing that event amplitudes in the laboratory scale in the same way as in Earth.

Our goal is to determine if the continuous seismic signal from the laboratory fault contains information about its frictional state. Recently, a seismic signal previously thought to be noise has been identified in the laboratory (\cite{Rouet2017}). This new signal has strong predictive ability regarding upcoming failures over the entire seismic cycle, suggesting that the seismic signal is imprinted with information about the fault frictional state.

\section{Machine learning finds the frictional state of the laboratory fault from the seismic signal it emits}

\begin{figure}[h!]
\begin{center}
\includegraphics[width=9cm,trim= 50 0 50 50]{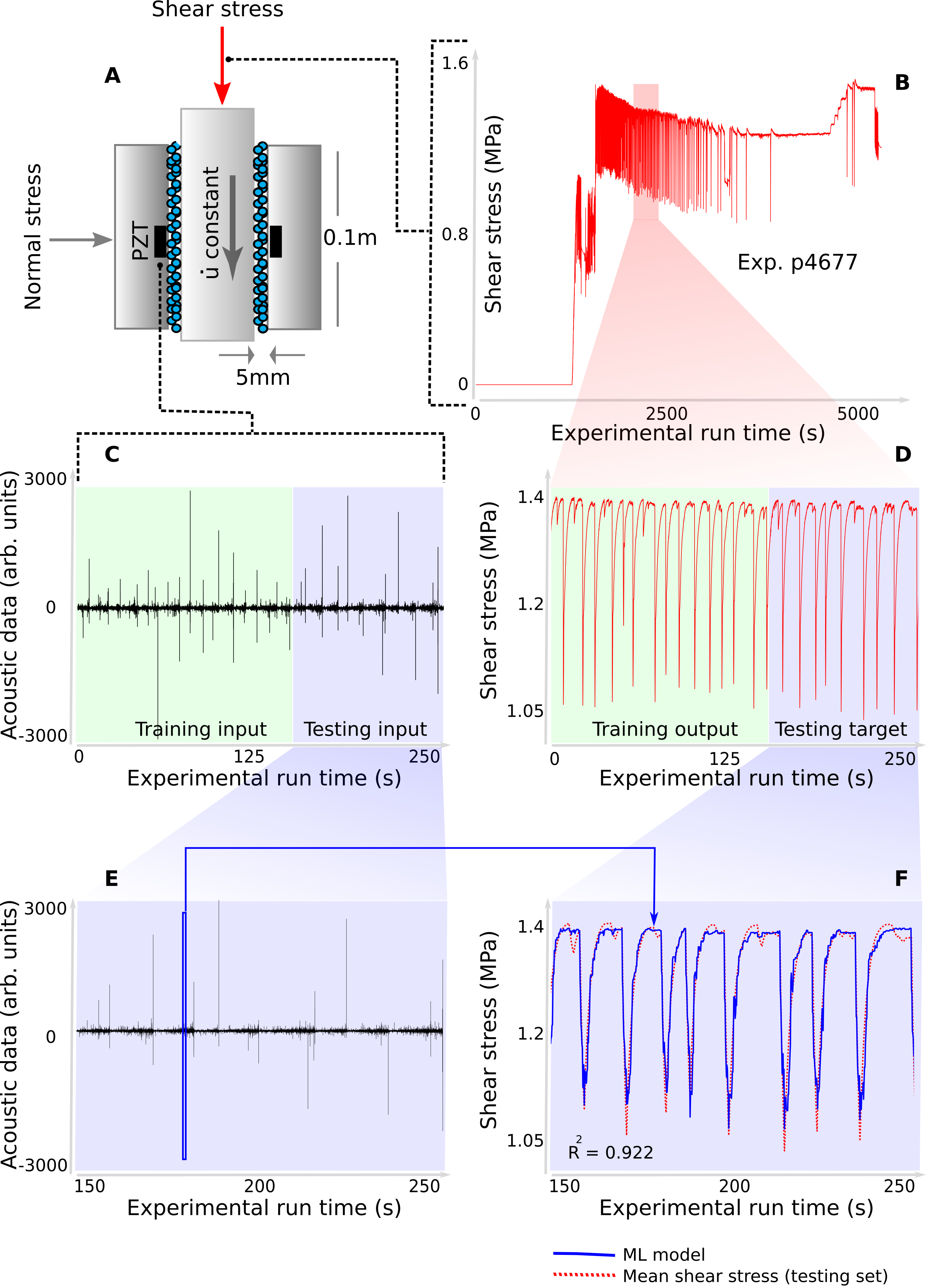} 
\caption{\textbf{The machine learning algorithm derives the stress on the laboratory fault from the seismic signal it emits.} (\textbf{A}) Experimental apparatus: bi-axial shearing of fault gouge under normal load (\cite{johnson2013,Kaproth2013}). (\textbf{B}) The shear stress during the full experiment. For our analysis we select a portion of the experiment (red shaded region) that exhibits aperiodic stick-slips (laboratory earthquakes). (\textbf{C}) Seismic signal recorded within the side blocks. (\textbf{D}) Shear stress recorded over the same time interval as (C). In both (C) and (D) the green shaded region corresponds to the training set, 60$\%$ of the data for which the algorithm has access to both the seismic data and the shear stress and tries to build a model relating the two. The blue shaded portion corresponds to the testing set, the remaining 40$\%$ of the data for which the algorithm has only access to the seismic data, and not the shear stress. The testing target in (D) is only used to evaluate the performance (R$^2$) of the model. (\textbf{E}) Seismic signal in the testing set. An example time window used in the ML analysis is drawn to scale in blue,  corresponding to a data point on the ML derived shear stress signal  shown in (F). (\textbf{F}) The  blue line is the shear stress derived by machine learning solely from the sequence of the small, overlapping moving time windows of the seismic signal. The dashed red line is the experimental shear stress data.}
\label{fig:stress_prediction}
\end{center}
\end{figure}

The experimental apparatus, a biaxial shear device, is a double direct shear device with an adjustable normal load (Fig. \ref{fig:stress_prediction}A).  A piston mimicking tectonic forcing drives a central block relative to two fixed side blocks. The two side blocks are separated from the central block by two layers of granular material, the fault gouge. The gouge layer thicknesses, shear stress, normal load and shear displacement are all recorded. The fault frictional state $\mu$, is given by the shear stress divided by the normal stress ($\mu=\sigma_{\rm S}/\sigma_{\rm N}$). In the following, we use shear stress and friction or frictional state interchangeably, as they are proportional at constant normal load. In the first experiment we analyze, the normal load is fixed at 2.5 MPa. In the second experiment, that we analyze in the next sections, the normal load is constant at load levels of 4, 5, 6, and 7 MPa.
The fault gouge is comprised of class IV glass beads with diameters 105-149 microns. The seismic signal (also known as acoustic emission) coming from the fault is recorded by piezoceramics embedded in the side blocks (see Methods for more details). The apparatus has been broadly discussed  in the literature (\cite{johnson2013,Kaproth2013,Scuderi2016,Marone1998}).

In order to study the fundamental friction physics of the fault system, we analyze the continuous seismic signal recorded during the experiment using a machine learning (ML) approach that is explicit and can thus be used to obtain physical information about the shear system. Our primary goal is to infer at all times the current frictional state of the fault, using information from short moving time windows of the seismic data (Fig. \ref{fig:stress_prediction}E, solid blue window). In each time window, we compute a set of potentially relevant statistical features that describe the distribution of the seismic signal. The ML model uses the features calculated in a time window to estimate the average shear stress (or friction) during that time window. The time windows we consider are 1.33 s in duration. The laboratory seismic cycle varies from 7 s to 17 s, with an average of $\approx$12s (Fig. \ref{fig:stress_prediction}D), and thus the time windows are snapshots of the instantaneous state of the fault system.

We used a ML algorithm known as gradient boosted trees (XGBoost implementation) (\cite{Chen2016,Friedman2000}), which is a decision tree ensemble method (\cite{Breiman1999}). The hyper-parameters of the gradient boosted trees model are determined using the EGO method (\cite{Jones1998,Rouet2016,Rouet2017a}), maximizing the performance in 5-fold cross-validation on the training set (see Methods for details). The training set, used to build the model, corresponds to the first $60\%$ of the experimental data, shown as the green shaded region in Fig. \ref{fig:stress_prediction}C and \ref{fig:stress_prediction}D. The testing set, used to evaluate the model's performance, corresponds to the remaining 40$\%$ of the data, shown as the blue shaded region in Fig. \ref{fig:stress_prediction}C and \ref{fig:stress_prediction}D.
Each decision tree estimates the frictional state using a sequence of decisions based on the statistical features derived from the time windows (see Methods). We train the gradient boosted trees model by providing the algorithm with both the time series of the measured friction and features of the measured seismic signal. We then test the resulting ML model on a portion of data not used in training (shown on Fig. \ref{fig:stress_prediction}, (E) and (F)). It is important to note that during the testing procedure, the ML model has access only to the features of the seismic data. In order to quantify the quality of the model's estimates of the frictional state compared to the experimental values, we use the coefficient of determination (R$^2$) as our evaluation metric.

Figure \ref{fig:stress_prediction}F shows that the ML model can accurately determine the instantaneous shear stress, \emph{i.e.} the frictional state, directly from instantaneous features of the seismic data. The statistical characteristics of any arbitrary segment of seismic data are a fingerprint of the associated fault frictional state. Despite the fact that the stress cycles are aperiodic, the ML model can determine the instantaneous frictional state of the fault from the seismic signal it emits, at all times. Importantly, the connection between instantaneous (local in time)  seismic features and instantaneous frictional state works throughout the entire seismic cycle.

In our experiments, the seismic signals come from grain fracture, rotation and displacement, or brittle failure of adhesive grain contact junctions within the laboratory fault gouge. Ongoing Discrete Element simulations (\cite{Dorostkar2017,Ferdowsi2015}) and Finite Element plus Discrete Element simulations are being applied to study the role of granular processes during shearing. 

Using machine learning, we showed that we are able to precisely infer the friction of a laboratory fault from statistical characteristics of the continuous seismic signal it emits. In the next section we will show that by probing the most important statistical feature identified in the seismic signal, we can extract a simpler model that does not have the same level of accuracy, but that is easier to interpret, can be generalized across experimental conditions, and from which we can uncover an equation of state linking fault friction and properties of the seismic signal.

\section{The laboratory fault exhibits a simple equation of state linking friction to seismic power, and exhibits a hysteretic behavior}

The frictional state determined by the ML model from the seismic data is highly accurate (R$^2 > 0.9$). A key characteristic of the ML decision tree models, and what makes them so valuable for the analysis of scientific data, is their simplicity and the fact that they are constructed explicitly from the features of the data they are provided with. This allows for a straight-forward ranking of the features based on their importance for the ML model (see Methods). Used in this way, the decision tree procedure enables us to determine which characteristic of the seismic signal is the most important to estimate the fault friction. Following this approach, we find that the key feature of the seismic signal is its variance. By definition, the variance of an elastic wave signal is proportional to the average energy per unit of time, thus it is proportional to the average power in the elastic wave signal during a time window. Therefore, it is straightforward to rebuild the frictional state ML model based solely on this single feature of the seismic signal. We show such a model, determined solely by the power in the seismic signal from the fault, in Figure 2.  Note that the estimated friction values $\mu$ remain accurate (R$^2 > 0.8$), which demonstrates a strong link between the power in the seismic signal from the fault and its frictional state.  Other features of the seismic signal are important, but less so than the variance (\emph{i.e.} seismic power). 

Fig. 2A shows the shear stress as a function of seismic power. The ML model built in training (where the ML model uses both the shear stress and seismic signals) is shown as a bold blue line.  The testing data are shown for the nine stress cycles (thin dashed lines) shown in Figure \ref{fig:stress_prediction}E and F. Fig. 2B shows several of the stick-slip cycles as a function of time, with colors corresponding to the data of Figure 2A.  The time window analysis  (e.g., see Figure 1E) used to construct the ML model of the frictional state during the training phase is established point-wise in time, over 1.33s intervals that are displaced by increments of 0.133s (90 percent overlap). Therefore, the model inputs contain no information about the timing of the failure events seen in Figure 2B.  The ML algorithm is able to estimate the frictional state, and therefore the position within the seismic cycle, based solely on the continuous seismic signal radiated by the fault.
Surprisingly, even though experimental shear stress trajectories differ in time, they are identical in seismic power-shear stress space. The training data can be scrambled in time and the frictional state model we find is unchanged. If we take the seismic signal to be generated by a spectrum of abrupt grain rearrangements driven by the shear stress, the relationship between shear stress or frictional state and the power in the seismic signal can be regarded as an equation of state.

\begin{figure}[h!]
\begin{center}
\includegraphics[width=12.5cm,trim= 50 0 50 20]{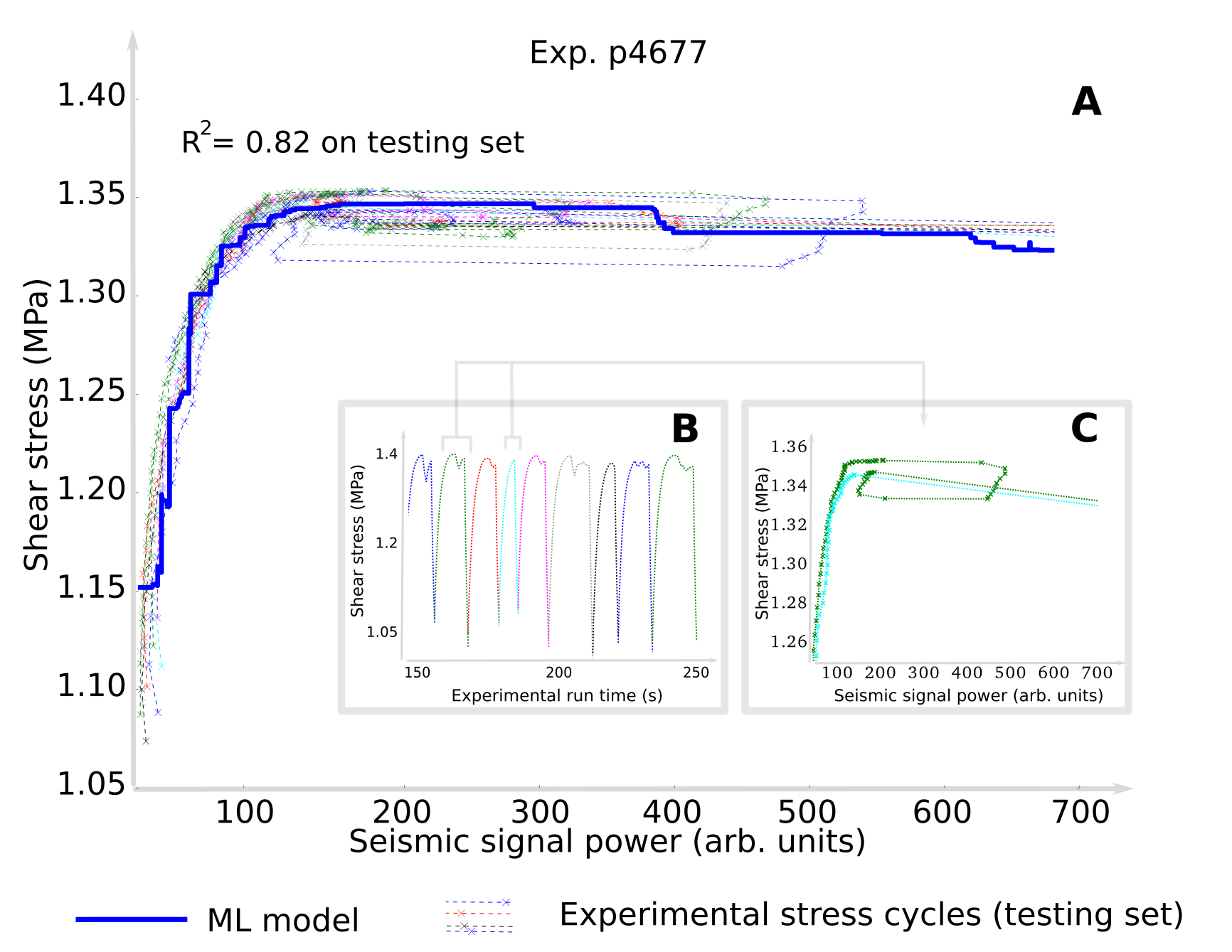}
\caption{\textbf{The laboratory fault exhibits a simple relationship between shear stress and seismic signal power.} (\textbf{A}) Thin dashed colored lines correspond to the experimental data (testing set), colored differently for each individual stick-slip cycle, shown in Fig. 2B. Each  data point (cross) represents the average shear stress on the fault and the measured seismic power obtained from a given time window (see Fig. 1E). Note the consistency of the individual shear stress-seismic power curves despite the differences in the stick-slip cycles in time (Fig. 2B). The bold blue line is obtained using \textit{both} the power of the seismic signal and the shear stress during the training procedure, on previous data. We emphasize that during testing here, the ML model sees only the seismic data, and not the shear stress data. The accuracy of the ML model during the testing procedure is remarkably good: From solely the instantaneous seismic power emitted by the fault, the ML model can make precise estimations of the stress for all individual stress cycles (colored dashed lines) with an R$^2$ of 0.82. (\textbf{B}) The laboratory stress cycles in time, with colors matching the stress cycles in (A). (\textbf{C}) Zoom of 2 laboratory stress cycles (second and fourth in (B) [note colors correspond]).  One of the two cycles  exhibits a distinct hysteresis loop due to a small shear failure preceding the primary failure, which are only sometimes observed.}
\label{fig:friction_law}
\end{center}
\end{figure}

Our results shown in Fig. 2A demonstrate a robust, predictive relationship between fault zone friction (or shear stress) and the power of the seismic signal coming from shear deformation within the fault gouge. This relation between seismic signal power and friction can be estimated by training the ML model on both seismic and shear stress data sets. This model is the bold blue line shown in Fig. 2A. In comparison, the thin dashed color lines in Fig. 2A and 2B come from the testing data that the ML model has never seen. The friction for any and all laboratory earthquake cycles can be calculated from this relation. Moreover, we find that this predictive relation holds for a broad range of conditions, including when the laboratory earthquake cycles are periodic, aperiodic and during the transient failure episodes as friction evolves (see Fig. S1 in the Methods section).  The results show that in the case of the laboratory fault, failure does not occur randomly, but on the contrary follows a very specific pattern given by an equation of state that links the friction on the fault to the power of the seismic signal it emits.

Interestingly, the laboratory seismic cycles show a complex behavior, with segments of quasi-steady stress prior to failure (Figure 2B).
During the critical stress state preceding failure the shear stress occasionally decreases, reflecting a small gouge failure, and then recovers (Figures 1D and 1F). This is manifested in a hysteresis loop in the friction vs. seismic power space (Figure 2C). The inset of Fig. 2 shows two stress cycles in seismic power-shear stress space, one with no inner loop (corresponding to no small stress drop during the cycle), and the other exhibiting an inner loop (corresponding to a small stress drop during the stress cycle). We draw a parallel between the hysterectic behavior that we find here and quasi-static experiments on rock (where  `discrete memory', also termed `end point memory' may occur when small stress cycles take place during a larger stress cycle (\cite{JGRB:JGRB3711})).

\FloatBarrier
\section{The equation of state linking friction to seismic power generalizes across load levels}

\begin{figure}[h!]
\begin{center}
\includegraphics[width=12.5cm,trim= 0 10 0 100]{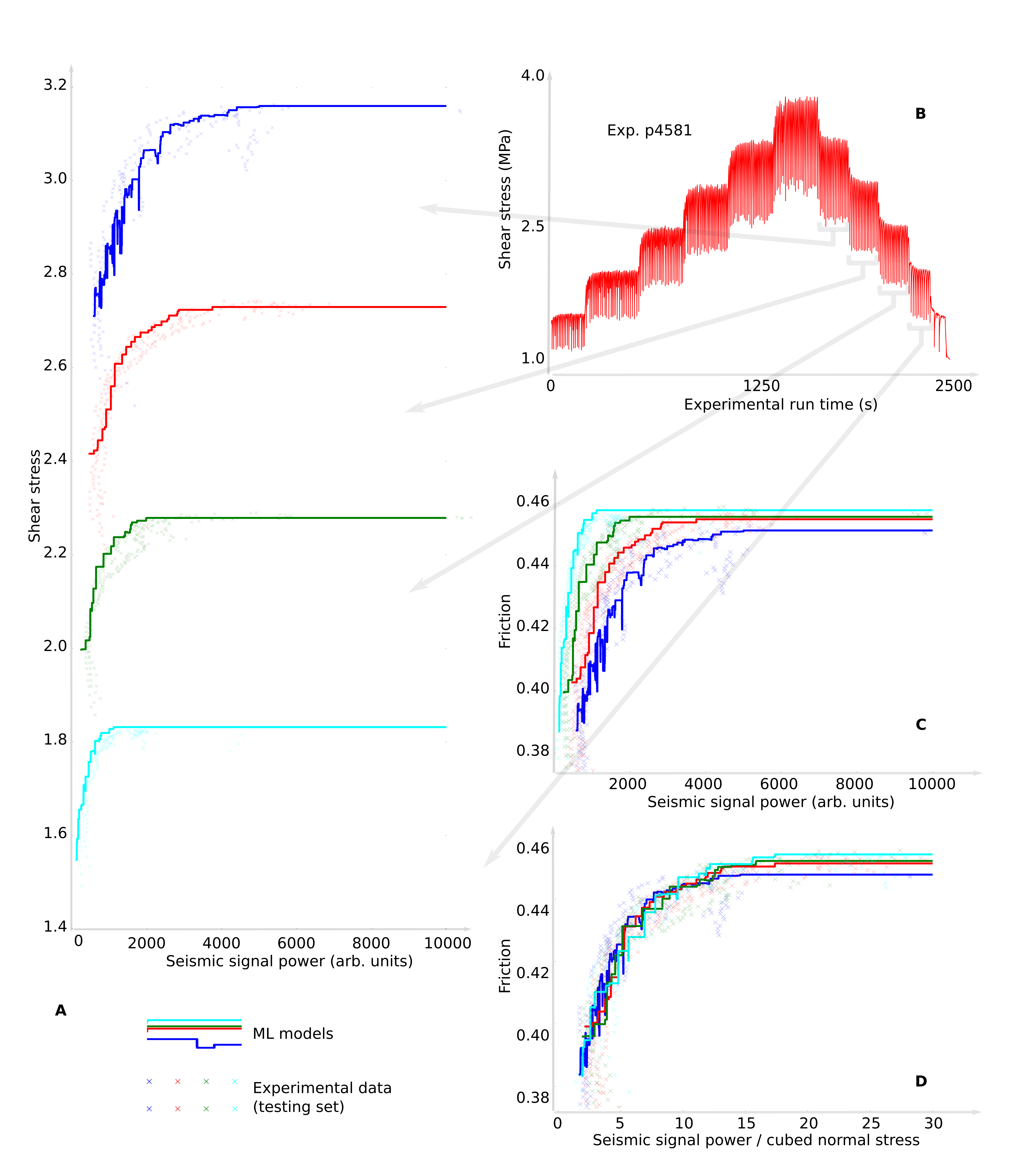}
\caption{\textbf{The machine learning-derived equation of state can be transferred across load levels.} (\textbf{A}) Seismic signal power-shear stress relations, for different normal loads (from bottom to top: 4, 5, 6, and 7 MPa). The thick solid lines are the respective relations built by machine learning on the training set, and the thin crosses are the laboratory data from the testing set. (\textbf{B}) The shear stress of the full experiment, noted with the gray arrows indicating the portion of the data we analyze. In contrast to the constant load experiment shown in Fig. 1, the first 80 percent of the data at each normal load is used for training, and the following 20 percent is used for testing. (\textbf{C}) The ML-derived friction laws shown on the same plot. The models are the same as in (A), with colors matching. The shear stress is normalized by the normal load to give the friction: friction = shear stress / normal stress. (\textbf{D}) The curves in (C), with seismic signal power normalized by the cube of the normal load.  All the friction laws collapse onto a single curve.}
\label{fig:normal_loads}
\end{center}
\end{figure}

The bi-axial apparatus enables us to study the laboratory seismic cycle for different normal loads (Fig. \ref{fig:normal_loads}B). In this section, we analyze a second experiment from the same apparatus, during which the normal load is progressively stepped up and then down. Equations of state similar to that in Fig.\ref{fig:normal_loads}A can be constructed for each normal load. The thick colored lines correspond to the equation of state linking friction or shear stress to seismic power estimated by the ML model for each load level (determined on the training set). The light colored crosses show the experimental trajectories the laboratory fault has gone through in seismic power-shear stress space (in the testing set, not used to build the ML model).  
In Fig. \ref{fig:normal_loads}C we show the estimated equation of state for each load level where seismic power is now plotted against frictional state instead of shear stress.  The different relations partially collapse onto one another. As a final step we scale the seismic power by the cube of the normal stress (Fig. \ref{fig:normal_loads}D). We find this scaling empirically from the data. A single \emph{universal} equation of state results.

The scaling of the equation of state linking friction to seismic power can be understood as arising from the properties of the fault gouge. The seismic signal is due to elastic waves broadcasts from the interior of the system that come from abrupt particle rearrangements. These rearrangements occur as the configurations of the granular material evolve to support larger and larger shear stress imposed by the drive. The granular material, modeled as a Hertzian material (\cite{Johnson1987}), involves particle-particle bonds that have energy, $e_{\rm B}$, that scales with the normal load as $e_{\rm B}\propto \sigma_{\rm N}^{5/3}$ (see Methods more details). We can assume that the elastic wave broadcasts that accompany rearrangements carry energy that scales as the bond energy $e_{\rm B}$. If we also assume that the set of particle configurations that unfold in a slip cycle are statistically the same for all values of the normal stress, then at a point in the slip cycle the elastic broadcasts will differ primarily due to the event rate $r$, as the slip cycle unfolds. Thus, the seismic power $P$ scales as: $P \propto e_{\rm B}r \propto \sigma_{\rm N}^{8/3} \approx \sigma_{\rm N}^{3}$.
To derive the third term, we use the observation that $r \propto \sigma_{\rm N}$: in the bi-ax experiment, the inter-event time is inversely proportional to the normal stress.

\begin{figure}[h!]
\begin{center}
\includegraphics[width=15cm,trim= 0 10 0 0]{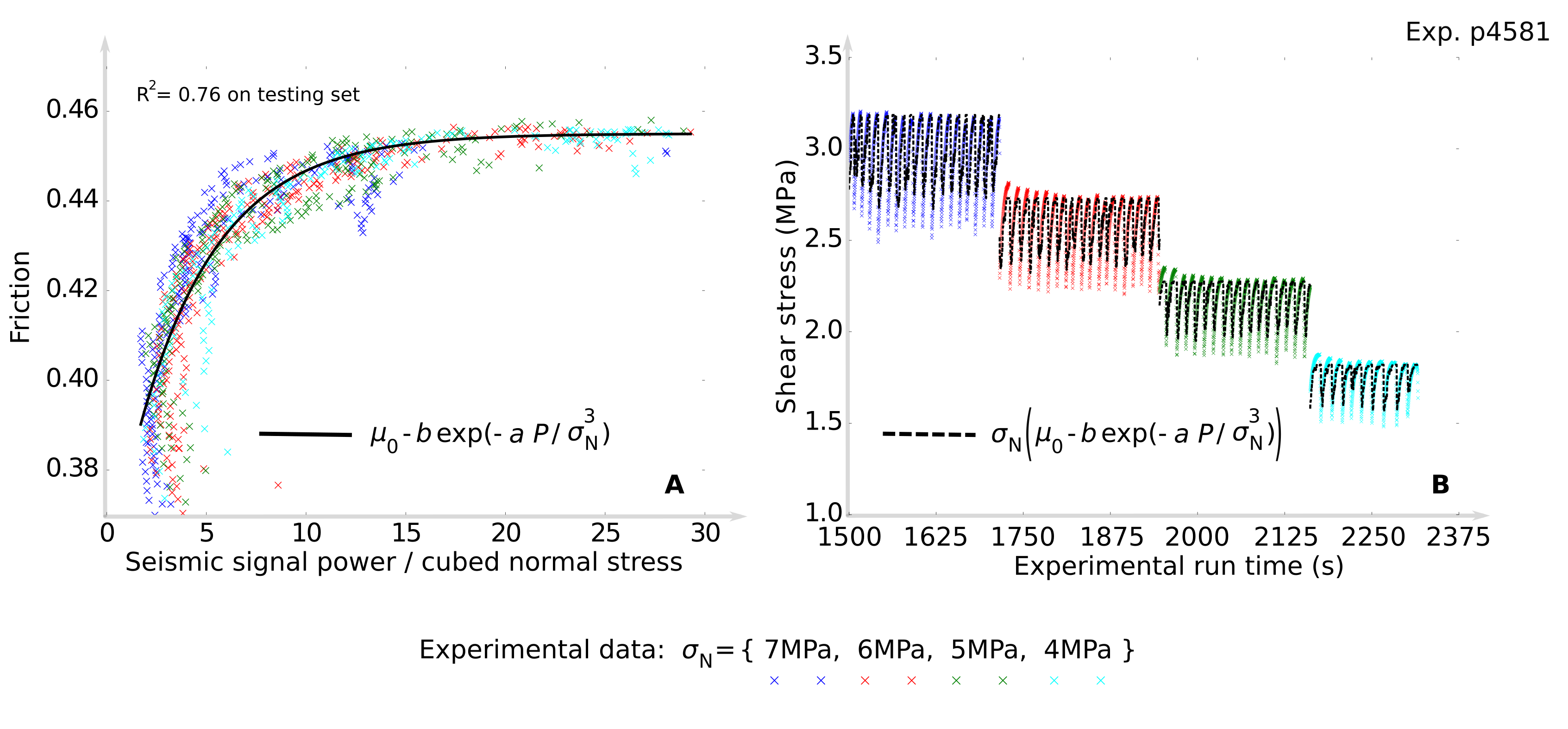}
\caption{\textbf{The friction on the fault follows a simple exponential function of the seismic signal power.} Once we know, thanks to the machine learning analysis, that the seismic signal power can accurately give the frictional state, we can use a rougher fit to visualize and interpret the seismic power-friction equation of state. The data shown are for the step-up step-down in load experiment.  (\textbf{A}) The thick black line corresponds to an exponential fit (Eq. 1). The fit is done only on the training set. The colored crosses are the same laboratory data from the testing set as on Fig. \ref{fig:normal_loads}, with color matching.
(\textbf{B}) Shear stress vs. time for a portion of the experiment.  The thick dashed line is the same exponential fit as in (A), converted to shear stress. At all times and displacements, the normalized seismic signal power gives a very accurate estimate of the shear stress (or friction), using the relation shown in black in (A). Here the colored crosses are the laboratory data for both training (on which the fit is done in seismic power-friction space) and testing. Given the power of the seismic signal within a small time window, the equation of state in Eq. 1 enables one to accurately determine the friction or shear stress of the fault at any moment of any stress cycle, at any load level. We note that the entire stress drop is correctly estimated using the ML model in Fig. 1, whereas only part of the stress strop is correctly estimated in (B) using Eq. 1.}
\label{fig:fit}
\end{center}
\end{figure}

The frictional state law derived by machine learning at one load level can therefore be transferred to any arbitrary load level by normalizing the seismic power by the cube of the normal stress. This  simple relation can give accurate estimations of the stress (or friction) on the fault for any stress cycle, at any load level. Moreover, once the machine learning analysis has established the direct relationship between seismic power and friction on the fault, we can use a simpler exponential fit to visualize this relationship. Such a simple fit is shown in Fig. \ref{fig:fit}E: 
\begin{equation}
\mu=\mu_0-b \exp \left(-a\frac{P}{\sigma^3_{\rm N}}\right) 
\end{equation}
\label{eq:fit}
with $\mu_0$ the asymptotic friction (reached at the end of the stress cycles and during stable sliding), $P$ the seismic power during a time window, $\sigma_{\rm N}$ the normal load, and $a=0.25$ and $b=0.1$ the parameters of the fit.

The laboratory fault assembly is opaque and therefore we cannot see inside to examine the behavior of the fault gouge. However, a simple interpretation of the seismic power we measure as coming from elastic energy stored in the granular material reproduces the scaling of the seismic power-friction law that we find. 

We have demonstrated that certain statistics of the seismic signal over short windows of time provide a fingerprint of the shear stress and frictional state of the fault. It is well established that failure in granular materials (\cite{Michlmayr2013}) is frequently accompanied by impulsive acoustic/seismic precursors. Precursors are also routinely observed soon before failure of a spectrum of industrial (\cite{Huang1998}) and Earth materials (\cite{Schubnel2013,jaeger2007poroelasticity}). Precursors are observed in laboratory faults (\cite{johnson2013,Goebel2013}) as well as models of faults (\cite{Daub2011,Latour2011}), and are widely but not systematically observed preceding earthquakes (\cite{bouchon2013,bouchon2016,McGuire2015,Mignan2014,Wyss1997,Geller1997}). The fingerprint that we find in the seismic signal emitted by the fault extends the observation of precursory seismic activity that often takes place soon before failure: we show that characteristics of the seismic signal can tell us about the frictional state of the laboratory fault not only right before failure, but at any time during the slip cycle.


\section{Conclusion} 
Our results show that the laboratory fault does not fail randomly but in a highly predictable manner. The observations also demonstrate that key  properties of the laboratory earthquake cycle can be inferred from the continuous seismic signal emitted by the fault. In particular, the instantaneous frictional state, the critical stress state and therefore where the fault is within the earthquake cycle can be determined using exclusively an equation of state that links the power of the continuous seismic signal to the friction on the fault. 
This tells us that at least in the laboratory, earthquake catalog approaches for analyzing fault physical characteristics are discarding critical information. Similar approaches using the continuous signal from seismic waves may yield new insight into faults in Earth.

\bibliographystyle{agujournal}

\subsection*{Acknowledgements}
We acknowledge funding from Institutional Support (LDRD) at Los Alamos National Laboratory, as well as funding from the US DOE Office of Fossil Energy. We thank Andrew Delorey, Kipton Barros, James Theiler, Marian Anghel, Jamal Mohd-Yusof and Nick Lubbers for helpful comments and/or discussions. All the data used are freely available on the data repository hosted by Chris Marone at the Pennsylvania State University.

\listofchanges

\end{document}